\magnification=\magstep1
\baselineskip=14pt
\overfullrule=0 pt
\font\titulobold=cmbx12 scaled\magstep1

\hfill IST-GDNL\#1/97

\hfill MIT-CTP\#2604

\hfill January~1997

\hfill {\tt hep-th/9704084}
\vskip 1 true cm
 
\titulobold
\centerline{Stable Knotted Strings} 

\tenrm
\bigskip
\centerline{\bf Rui Dil\~ao$^a$\footnote\dag{\rm rui@sd.ist.utl.pt}\ \ {\tenrm and}\ \ Ricardo Schiappa$^{b}\footnote\ddag{\rm ricardos@mit.edu}$}
\bigskip
\centerline{\it a-  Grupo de Din\^amica N\~ao-Linear, Departamento de F\'\i sica}
\centerline{\it Instituto Superior T\'ecnico, Av. Rovisco Pais}
\centerline{\it 1096 Lisboa Codex, Portugal}
\medskip
\centerline{\it b-  Center for Theoretical Physics and Department of Physics}
\centerline{\it Massachusetts Institute of Technology, 77 Massachusetts Av.}
\centerline{\it Cambridge, MA 02139, U.S.A} 

\bigskip
\bigskip

\centerline{\bf Abstract}

We solve the Cauchy problem for the relativistic  closed string 
in Minkowski space $M^{3+1}$, including the cases where the initial data has a
knot like topology. 
We give the general conditions for the world sheet of a closed
knotted string to be a time periodic surface. In the particular case of zero
initial string velocity the period of the world sheet is proportional to half
the length ($\ell$) of the initial string and a knotted string always
collapses to a link for $t=\ell/4$.
Relativistic closed strings are dynamically evolving or pulsating structures in spacetime, and knotted or 
unknotted like structures remain stable over time. 
The generation of arbitrary $n$-fold knots,
starting with an initial simple link configuration with non zero
velocity is possible. 
 
\bigskip
\bigskip
\centerline{\sl Phys. Lett. B {\bf 404} (1997) 57-65}

\vfill


\eject
\bigskip
\bigskip
\bigskip

{\bf 1. Introduction}

\medskip

String theories and knot theory are playing a prominent role in modern theoretical physics, [1] and [2]. The foundations go  back 
to Thomson's recognition of the invariance of vortex tubes in incompressible and non-viscous fluid flows and the hypothesis that atoms were knotted vortex
tubes in ether [3].
In modern physics, string  theory is becoming an important field of research
due to the rich phenomena and physical concepts associated to nonlinear 
field theories in superstring theories [1], cosmic string theory [4-5], 
statistical mechanics [6], molecular biology [7] and $^3$He superfluids [8].

In the thirties Born and Infeld [9] proposed that the natural laws are
independent of the spacetime coordinate systems and this should be expressed
as the invariance of the action for any change of parameterization.
Later, Nambu [10] proposed a relativistic string theory based on the 
invariance of the action for any parameterization of the world sheet of a 
string in Minkowski space. Therefore, 
the two-dimensional scalar equation of the Born-Infeld nonlinear
electrodynamics  is the equation of motion of a string in the Minkowski
space $M^{2+1}$, [11], and the solution of this equation defines a
surface of minimal area in   $M^{2+1}$, the world sheet of a string.   

The Cauchy problem for the equation of motion of an infinite relativistic
string in $M^{2+1}$ has been solved [11].
The nonlinear nature of the solutions of this equation 
leads to the appearance of soliton like propagation,
characterized by the coherence   of wave packets before and after 
interaction, [12]. However, singularities arise when 
large amplitude  travelling wave solutions  collide [13]. 
Geometrically, these singularities correspond to the existence of folds and
cusps in the integral surface associated to the solutions of the 
relativistic string equation, [12].

On the context of galaxy formation, Zeldovich  suggested that strings formed
in the early universe might provide the density perturbations needed to start the
condensation process. Different grand unification phase transition temperatures derive from the string lifetimes and lengths. Kibble and
Turok [4] have shown that any initially static closed string can self-intersect and collapse to a double loop, and conjectured that they can annihilate into particles. Hindmarsh and Kibble [5] have shown that for zero initial 
string velocity the world sheet of the string is a time periodic surface with
period  proportional to half the length of the initial string.

Besides the classical studies of solitons in strings, there has been a 
growing interest in string theory in the last few years, in the framework of
quantum fields. In fact, the quantum study of solitonic solutions in superstring theory has become of
central importance in modern string theory, as a nonperturbative probe into
the strong coupling regime [14]. 

The Cauchy problem for open string configurations (free end points) 
appeared in the literature, [11], [15],  [16] and [1]. 
The classical and quantum mechanics of a massless relativistic string with
free end points moving at the speed of light has been analysed in   [15].
However, as pointed out by Faddeev and Niemi [17], no explicit analysis of 
stable knotted structures exists for nonlinear field equations. 

Here we solve the general Cauchy
problem for the closed knotted string in $M^{3+1}$,
showing that propagating 
knot like structures in Minkowski space $M^{3+1}$ are preserved, generating a
time periodic world sheet.
The singularity theory associated to string loop annihilation and creation
is developed.
In particular, we show that initially closed static strings always oscillate 
with the period $T=\ell /2$, where $\ell $ is the string length.  
Knotted strings evolve to simple links with crossover, and  initially
closed simple links collapse  to  points. From points  and simple links
emerge closed knotted strings. 
The temporal periodicity of
the string implies an infinite lifetime with periodic transitions from the simple link to the knotted string. We also show the existence of finite length
string configurations  that stretch indefinitely. Our analytical results
are supplemented with simulations of the time evolution of the simple
link, trefoil and figure eight knot.

\bigskip

{\bf 2. The Cauchy Problem for Knotted Strings}

\medskip

We consider the Minkowsky space $M^{3+1}$  with metric
$d\ell^2=dt^2-dx^2-dy^2-dz^2=(dx^0)^2-(dx^1)^2-(dx^2)^2-(dx^3)^2$, and units
such that $c=1$.
A string $C$ in a general position in space at time $x^0=t$ is described by
the set of spatial coordinates
$$
x^{k}(t,\sigma) \, , \quad k=1,2,3 \eqno(2.1)
$$
where $\sigma $ is a parameter and $t$   
is measured in the Lorentz frame of the string. We call $t$ the proper time 
of the string.

A propagating string $C$ in $M^{3+1}$ defines  a two-dimensional surface $S$ 
with boundary $\partial S=C$, the world sheet of the string. The intersection of $S$ with the hyperplane $x^0=t$ is the configuration of the
string at time $t$.

In general, in a metric space, the first fundamental form of a surface
is a quadratic form $g_{ij}dx^idx^j$, and
the infinitesimal area of a surface  is proportional 
to the square root of the determinant of the covariant 
tensor $g_{ij}$. Nambu [10] 
postulated that the action for the motion of the string in Minkowski space is proportional to
the area of its world sheet $S$,   
$$
A=-m\int\int dp\ dq\ \sqrt{
(\partial_p {\vec r}\cdot \partial_q {\vec r})^2 -
(\partial_p {\vec r})^2 (\partial_q {\vec r})^2}\eqno(2.2)
$$
where ${\vec r}=(x^0(p,q),x^1(p,q),x^2(p,q),x^3(p,q))$,
$p$ and $q$ are the parameters of the surface $S$ and $m$ is a constant of
proportionality with dimension of (length)$^{-2}$. 
As the area of a surface, $\int \int \sqrt{g} dpdq$, is independent of the parameterization, it is proportional to the action (2.2).

For a string in $M^{3+1}$ it is natural to take $p=t$ and $q=\sigma$, but  
in general $p$ and $q$ are functions of $t$ and $\sigma $. In the following we adopt this more general viewpoint.

The Euler-Lagrange equations for the motion of the string in spacetime are 
derived from the  action (2.2), where the Lagrangian is 
${\cal L}=-m\sqrt{
(\partial_p {\vec r}\cdot \partial_q {\vec r})^2 -
(\partial_p {\vec r})^2 (\partial_q {\vec r})^2}$. 
However, with the choice $p=t$ and $q=\sigma$ we obtain  three second 
order nonlinear 
partial differential equations that are difficult to analyse. Instead, we
take the arbitrary parameterization $p=p(t,\sigma )$ and $q=q(t,\sigma )$ and
we  simplify the action with a trick introduced by
Barbashov and Chernikov [11]. These authors have shown that the hyperbolicity of the Euler-Lagrange equations implies a choice of the parameters $p$ and $q$ in such a way that
$$
(\partial_p {\vec r})^2=0\, ,\quad (\partial_q {\vec r})^2=0\eqno(2.3)
$$
Under these conditions, the   action (2.2) simplifies,
$$
A=-m\int\int dp\ dq\   (\partial_p {\vec r}\cdot \partial_q {\vec r}) 
\eqno(2.4)
$$
With the new parameters $p$ and $q$   obeying the
constraint equations (2.3) and to be determined later,
the Euler-Lagrange equation of motion for  the free string in Minkowski
space is
$$
\partial_p \partial_q {\vec r}=0\eqno(2.5)
$$

Equation (2.5) is readily solved and the general solution is
$$
x^{\mu}(p,q)=f^{\mu}(p)+g^{\mu}(q)\, , \quad \mu =0,1,2,3  \eqno(2.6)
$$
where $f^{\mu}$ and $g^{\mu}$ are arbitrary functions that must be determined from initial data and the constraint equations (2.3).
In order to determine  $f^{\mu}$ and $g^{\mu}$ from
initial data and  constraint equations --- Cauchy problem ---, we   express the solutions (2.6) in the 
form (2.1),
$$
x^{\mu}(t,\sigma)=x^{\mu}(p(t,\sigma),q(t,\sigma))\, , \quad \mu=0,1,2,3\eqno(2.7)
$$ 

At $t=0$, the initial position and velocity of a string is specified parametrically by the functions
$$\eqalign{
x^{k}(0,\sigma)&=a^{k}(\sigma)\cr
\partial_tx^{k}(t,\sigma)|_{t=0}&=b^{k}(\sigma)\, , \quad k=1,2,3\cr}\eqno(2.8)
$$
If the $a^{k}(\sigma)$ and $b^{k}(\sigma)$
are periodic functions in $\sigma$, these functions
describe a  closed curve (knotted or unknotted) in the subspace $t=0$ of 
$M^{3+1}$. Without loss of generality, we take the periodicity conditions
$a^{k}(\sigma)=a^{k}(\sigma+2\pi )$ and $b^{k}(\sigma)=b^{k}(\sigma+2\pi )$,
where $2\pi$ is the least period for the functions $a^{k}(\sigma)$.  
Initial string velocities can have least period $2\pi/n$, where $n$
is some positive integer.

As the parameters $p$ and $q$ are arbitrary,  the constraint equations (2.3) do
not change if we take $p=h_1(p')$ and $q=h_2(q')$, with $(\partial_{p'} h_1)\not=0$
and $(\partial_{q'} h_2)\not=0$. Therefore, for $t=0$, we choose 
$p=q=\sigma$ ($p'=q'=\sigma$, and $h_1$ and $h_2$ are the identity, for $t=0$). Under these conditions,  
comparing (2.8) with (2.6), we have, for $t=0$,
$$\eqalign{
x^0(p,q)|_{p=q}&=t(p,q)|_{p=q}=f^0(\sigma)+g^0(\sigma)=0\cr
x^k(p,q)|_{p=q}&=f^k(\sigma)+g^k(\sigma)=a^k(\sigma)\, ,\quad k=1,2,3\cr
 }\eqno(2.9)
$$

Writing the solution (2.6) in d'Alembertian form
$$
{\vec r}(p,q)={1\over 2} ({\vec \rho } (p)+{\vec \rho}(q))+
{1\over 2}\int_p^q {\vec \pi }(s)ds
\eqno(2.10)
$$
and comparing (2.10) with (2.9) for $t=0$, we conclude that
${\vec \rho }(p)|_{t=0}=(0,a^1(\sigma),a^2(\sigma),a^3(\sigma))$. So, the solution 
(2.6) can now be written as
$$\eqalign{
t(p,q)&= {1\over 2}\int_p^q \pi^0(s)ds\cr
x^k(p,q)&={1\over 2} (a^k(p)+a^k(q))+{1\over 2}\int_p^q \pi^k(s)ds
\, ,\quad k=1,2,3\cr}
\eqno(2.11)
$$
where $\pi^{\mu}$ are four unknown functions. Calculating the derivatives
$\partial_p x^k$ and $\partial_q x^k$ from (2.11), with $p$ and $q$ as functions of $t$ and $\sigma$, and solving for $\partial_t x^k$ and 
$\partial_{\sigma } x^k$ at $t=0$, $p=q$, $\partial_p \sigma |_{t=0}  =\partial_q \sigma |_{t=0}$, and
$\partial_t x^k|_{t=0}=\pi^k(\sigma)/\pi^0(\sigma)=b^k (\sigma)$. 
Therefore, $\pi^k=b^k \pi^0$ are directly computed from initial velocities 
(2.8) and $\pi^0$. For $t=0$,  the constraint equations (2.3) become
$$\eqalign{
(\pi^0)^2&=(\partial_{\sigma} a^1-b^1 \pi^0)^2+
(\partial_{\sigma}a^2-b^2 \pi^0)^2+
(\partial_{\sigma} a^3-b^3 \pi^0)^2\cr
(\pi^0)^2&=(\partial_{\sigma} a^1+b^1 \pi^0)^2+
(\partial_{\sigma}a^2+b^2 \pi^0)^2+
(\partial_{\sigma} a^3+b^3 \pi^0)^2\cr
}
\eqno(2.12)
$$
After developing the right hand side of (2.12), we finally obtain,
$$\eqalign{
(\pi^0 )^2\left(1-\sum_{i=1}^3(b^i )^2\right)+
2\pi^0 \left( \sum_{i=1}^3 b^i  {\partial_{\sigma}a^i } \right)-
\left( \sum_{i=1}^3({\partial_{\sigma}a^i })^2\right)&=0\cr
(\pi^0 )^2\left(1-\sum_{i=1}^3(b^i )^2\right)-
2\pi^0 \left( \sum_{i=1}^3 b^i  {\partial_{\sigma}a^i } \right)-
\left( \sum_{i=1}^3({\partial_{\sigma}a^i })^2\right)&=0\cr\cr
}
\eqno(2.13)
$$

Solving the above equations, we obtain the following solution, 
compatible with the $p,q$-parameterization: 
$$
\pi^0 (\sigma)= \pm\left({ {\sum_{i=1}^3 (\partial_{\sigma} a^i(\sigma))^2} 
\over
1-\sum_{i=1}^3(b^i (\sigma))^2
}\right)^{1/2}\eqno(2.14)
$$
provided   $0\le \sum_{i=1}^3(b^i (\sigma))^2<1$ and 
$\sum_{i=1}^3 b^i (\sigma) {\partial_{\sigma}a^i(\sigma)}=0$.

With, $\pi^k(\sigma)=b^k (\sigma)\pi^0(\sigma)$ and
introducing (2.14) into (2.11),  the   forward time solution 
of equation (2.5) is 
$$\eqalign{
t(p,q)&=  {1\over 2}\int_p^q  \left({ 
{\sum_{i=1}^3 (\partial_{s} a^i(s))^2} \over
1-\sum_{i=1}^3(b^i (s))^2
}\right)^{1/2}ds\cr
x^k(p,q)&={1\over 2} (a^k(p)+a^k(q))+
{1\over 2}\int_p^q  b^k(s)\left({ 
{\sum_{i=1}^3 (\partial_{s} a^i(s))^2} \over
1-\sum_{i=1}^3(b^i (s))^2
}\right)^{1/2}ds
\, ,\   k=1,2,3\cr}\eqno(2.15)
$$
provided $0\le \sum_{i=1}^3(b^i (\sigma))^2<1$, 
$\sum_{i=1}^3 b^i (\sigma) {\partial_{\sigma}a^i(\sigma)}=0$ and $p=q=\sigma$,
for $t=0$. In the particular case of zero initial velocities, $b^k (\sigma)=0$,
for $k=1,2,3$, we have
$$\eqalignno{
t(p,q)&=  {1\over 2}\int_p^q  \left({ 
{\sum_{i=1}^3 (\partial_{s} a^i(s))^2}  
}\right)^{1/2}ds& (2.16a)\cr 
x^k(p,q)&={1\over 2} (a^k(p)+a^k(q))
\, ,\   k=1,2,3& (2.16b)\cr } 
$$
Note that, if we take   $q\ge p$, the choice of the plus sign in (2.14) is equivalent to say that the solution (2.15) is forward in time. If we 
let $q<p$ in (2.15), we obtain a solution backwards in time, that could also be
calculated with the choice of the minus sign in (2.14) with $q\ge p$. Therefore,
with $p,q\in {\bf R}$, the choice of the plus or minus sign in (2.14) leads to the same solution (2.15). It can be straightforwardly shown that this reversibility property of the solutions (2.15) are derived from the nonlinear equations of motion 
in the $t$ and $\sigma $ coordinates, directly derived from the 
action (2.2)  with $p=t$ and $q=\sigma$.  

The condition  $0\le \sum_{i=1}^3(b^i (\sigma))^2<1$ on the initial 
data   imposes the constraint that the transverse string velocity is always
below $1$. (Note that according to our choice of units $c=1$).
The second condition, $\sum_{i=1}^3 b^i (\sigma) {\partial_{\sigma}a^i(\sigma)}= 0$, is a transversal  constraint between
the initial velocity and the parametric representation of the string, and
can be written in the form $(\partial_t \vec r \cdot \partial_\sigma \vec r)|_{t=0}=0$. Physically, this means that the solution (2.16) 
exists as far as strings are not allowed to break.

Hence, the Cauchy problem for a closed knotted string in $M^{3+1}$ is solved.

From the above solution we derive now an important property for the 
topology of the world sheet of the string. Let a closed string 
be parameterized at $t=0$ by periodic functions $a^k(\sigma)$, with
$k=1,2,3$, and $a^k(\sigma +2\pi)=a^k(\sigma)$, with $\sigma \in {\bf R}$.
By construction, $b^k(\sigma)$ is also periodic with period $2\pi /n$,
where $n$ is a positive integer.
By (2.15), it follows that 
$$
x^k(p,q+2\pi)=x^k(p,q)+{1\over 2}\int_q^{q+2\pi}  b^k(s)\left({ 
{\sum_{i=1}^3 (\partial_{s} a^i(s))^2} \over
1-\sum_{i=1}^3(b^i (s))^2
}\right)^{1/2}ds
\, ,\   k=1,2,3\eqno(2.17)
$$
As the integrand in the above equation has least period
$2\pi$, and if the functions $b^k(\sigma)$ are such that the three integrals 
in (2.17) are zero, we have $x^k(p,q+2\pi)=x^k(p,q)$ and, therefore,  
the world sheet $S$ is periodic in the spatial coordinates $x^k$.   
To determine the proper time period of the world sheet, we have,  by (2.15),
$$
t(p,q+2\pi)=t(p,q)+{1\over 2}\int_q^{q+2\pi}  \left({ 
{\sum_{i=1}^3 (\partial_{s} a^i(s))^2} \over
1-\sum_{i=1}^3(b^i (s))^2
}\right)^{1/2}ds:=t(p,q)+T\eqno(2.18)
$$
where $T$ is the period of the world sheet. In the particular case where
$b^k(\sigma )=0$, for $k=1,2,3$, we obtain $T=\ell /2$, where 
$\ell $ is the length of the string at  $t=0$, [5].
 (Note that $c=1$. In SI units
we have $T=\ell /(2c)$). So,  zero initial velocity closed strings  pulsate
in  spacetime. If some $b^i(\sigma) \not= 0$ and the integral in (2.17) 
is zero, the period is given by the integral in (2.18).

If the integral in (2.17) is non zero, then, this integral as a function of
$q$, will be unbounded with increasing $q$ and the  world sheet of the string 
is no longer periodic in time. As $t(p,q)$ is an increasing function of $q$,
in the limit $t\to \infty$, $x^k\to \pm \infty$.
In this case, the string will become infinitely long as proper time passes.

\bigskip

{\bf 3. Singularities and Knotted String Solutions}

\medskip

To depict the configuration of a string in $M^{3+1}$ at time $t$, 
with zero initial velocities, 
we  solve the implicit equation (2.16a). So,  for fixed
$p=p_0$ and $t=t_0$, the value of $q$ is the root of the equation
$$
{1\over 2}\int_{p_0}^q  \left(  
{\sum_{i=1}^3 (\partial_{s} a^i(s))^2} \right)^{1/2}ds-t_0=0\eqno(3.1)
$$
Let $q_0$ be a root of (3.1).
Therefore, if $q_{00}$ is an initial guess for the value of $q_0$, by the Newton method,
$$
q_{0n+1}=q_{0n}-2{t(p_0,q_{0n})-t_0\over 
\left(  \sum_{i=1}^3 (\partial_{s} a^i(s)|_{s=q_{0n}})^2\right)^{1/2}}
\eqno(3.2)
$$ 
and in the limit $n\to \infty $, $q_{0n}\to q_0$. Therefore, for a given
$t_0$ and $p_0$ we determine $q_0$ by (3.2), and the string configuration
is given by (2.16b). The general case with $b^k\not=0$ is analogously derived
from (2.15).

In Fig. 1,  we show the world sheet of the simple link 
in the Minkowski space $M^{2+1}$, for  zero initial velocity.  
In Figs. 2 and 3, the trefoil and figure eight knot configurations at successive times are shown, for initially static strings.
In these cases, the world sheets are time
periodic surfaces, as seen in \S 2. However, in these simulations,
there exists a time $t=t'$, with $0<t'<T$, for which the string loop
self intersects, leading to the
loss of injectivity of the map $(p,q)\to S\in M^{3+1}$, defined by (2.15).
Under these circumstances, we say that the string develops  a singularity.
Also, this singularity depends on the type of knot we choose as initial
configuration. As seen in Fig. 1, 2 and 3, the simple link collapses to a point, 
while both the trefoil and figure eight knot  collapse to a simple link.

In order to analyse the string singularities during time evolution, we introduce the
derivative $D$ of the map $(p,q)\to S$.
By  (2.11),
$$
D=\pmatrix{\partial_p t  & \partial_q t\cr
\partial_p x^1  & \partial_q x^1 \cr
\partial_p x^2  & \partial_q x^2 \cr
\partial_p x^3  & \partial_q x^3 \cr
}={1\over 2}\pmatrix{ -\pi^0(p)&\pi^0(q)\cr
\partial_p a^1(p)-\pi^1(p)&\partial_q a^1(q)+\pi^1(q)\cr
\partial_p a^2(p)-\pi^2(p)&\partial_q a^2(q)+\pi^2(q)\cr
\partial_p a^3(p)-\pi^3(p)&\partial_q a^3(q)+\pi^3(q)\cr
}
$$ 
and the point $\vec r(p,q)\in S$ is nonsingular if the rank of $D$ is 2.
The singularities on the world sheet $S$  occur for  $p$ and $q$ values
where the rank of $D$ is 1. So,
as the row and column ranks of a matrix are equal, the matrix D has rank
1 if any two  row vectors of $D$ are linearly dependent. Therefore, let $\alpha$ and $\beta$ be constants such that $\alpha(-\pi^0(p),\pi^0(q))+\beta
(\partial_p a^k(p)-\pi^k(p),\partial_q a^k(q)+\pi^k(q))=0$. Solving this 
equation for $\alpha$ and $\beta$, it follows that any two row vectors of $D$
are linearly dependent if $p$ and $q$ are simultaneous solutions of the equations 
$$
\pi^0(q)(\partial_p a^k(p)-\pi^k(p))+\pi^0(p)(\partial_q a^k(q)+\pi^k(q))=0\,
,\ k=1,2,3 \eqno(3.3)
$$

Let us now analyse the singularities of the simple link,  trefoil and figure eight knot, 
in the cases of Figs. 1, 2 and 3.

For the simple link of Fig. 1, we have  $a^1(\sigma)=\cos(\sigma)$,
$a^2(\sigma)=\sin(\sigma)$ and $b^k(\sigma)=0$, for $k=1,2$. 
As $\pi^k=0$, for $k=1,2$, by (3.3),
the singular values of the derivative $D$, in the $p$ and $q$ plane,
is the set of points $(p,q)$ that are simultaneous solutions
of the equations $-\sin(p)-\sin(q)=0$ and $\cos(p)+\cos(q)=0$. Hence,
these singular values lie 
on the graphs of the curves $q=p\pm \pi+2 \pi n$ in the $(p,q)$ plane,
where $n$ ranges over all the integers.
Introducing the relation $q=p\pm \pi+2 \pi n$ into (2.16a), for
this particular case of initial data, it follows that singularities 
do occur for $t=(2n+1)\ell/4$, where $\ell=2\pi$ is the length of
the string at time $t=0$, Fig. 1. Introducing 
$(p,q)=(p,p\pm \pi+2 \pi n)$ into (2.16b), we obtain $x^k(p,q)=0$ for all 
$p,q\in {\bf R}$. Therefore,    for $t=(2n+1)\ell/4$, the string collapses 
into a point. Outside these singularities, the simple link configuration is stable, evolving in spacetime $M^{2+1}$ as a pulsating structure.

The singularity analysis for the trefoil knot is more evolved.
In Fig. 2 we took the initial parameterization 
$a^1(\sigma)=(2.0+0.6\cos(3\sigma))\cos(2 \sigma)$, 
$a^2(\sigma)=(2.0+0.6\cos(3\sigma))\sin(2 \sigma)$,
$a^3(\sigma)=2 \sin(3 \sigma)$ 
and $b^k(\sigma)=0$, for $k=1,2,3$.   Singular solutions
of the equations of motion occur at the 
triple intersections of the zero level set of the three functions in (3.3),
Fig. 4a). In this case, numerical analysis shows that the singularity occurs
for $t=\ell /4$, where, in this example, $\ell=37.1587$. At the singular points
the trefoil knot configuration of the string collapses to a simple link. 
Outside the singularities, the trefoil knot is stable. 

In the figure eight knot, Fig. 3, the periodicity of the world sheet is also $T=\ell /2$, but
singular solutions occur for several proper time values between $0$
and $\ell /2$. In this parameterization, in each half period
of the world sheet there exist 
several time values for which the knot folds cross each other,
unknotting and knotting the closed string during time evolution.
 In this case, the
zero level sets of the three functions in (3.3) are presented in Fig. 4b).
Observe however that for $t=\ell /4=4.45423$, the singular solution  
corresponds to the simple link. Outside
these singularities, the figure eight knot is stable.

The existence of singularities at $t=\ell/4$ in the examples of Fig. 1-3,
lead to the conjecture that every knot configuration of an initially static string will cross the simple link configuration, independently of the initial parameterization. 
This $t=\ell/4$ collapse to a link can then be seen as a universal property
for knotted strings. 

Due to the reversibility property of the solution (2.15), 
we can have a simple link as initial configuration and obtain a $n$-fold knot.  
This is easily obtained with nonzero initial velocities.

\bigskip

{\bf 4. Conclusions}

\medskip

We have solved explicitly the Cauchy problem for closed knotted strings in
Minkowski space $M^{3+1}$.  We have shown that the world sheet of a 
closed string can be periodic in time or can stretch infinitely as $t\to\infty$.
The period has been calculated and is related with the string dimensions. 
Strings are always stable structures in spacetime. Due to the
reversibility property of the equations of motion,  the dynamical
generation of $n$-fold  knotted strings from simple links is possible.

Future work in exact solutions of string theories include the study of 
(self-gravitating)  knotted strings in a Friedman-Robertson-Walker spacetime and quantization
of knotted strings. Important questions arising from this work
concern the effect of singularities on string diagram calculus, [1],
and changes in the partition function due to knotted
strings [18].

\bigskip

{\bf Acknowledgments:} We would like to thank Kenneth Johnson for a careful
reading of the manuscript. One of us (RS) is partially supported by the
Praxis XXI grant BD-3372/94 (Portugal).

\vfill \eject
\centerline{\bf References}
\bigskip 

[1] --- M. B. Green, J. H. Schwarz \& E. Witten, {\it Superstring Theory}, 
Cambridge Univ. Press (1987). 

[2] --- L. H. Kauffman, {\it Knots and Physics}, World-Scientific (1991). 

[3] --- W. Thomson, {\it On Vortex Motion}, Trans. R. Soc. Edinburgh 
{\bf XXV} (1869) 217-260. 

[4] --- T. W. B. Kibble \& N. Turok {\it Self-Intersection of Cosmic
Strings}, Phys. Lett. {\bf 116B} (1982) 141-143. 

[5] --- M. B. Hindmarsh \& T. W. B. Kibble, {\it Cosmic Strings}, 
Rep. Prog. Phys. {\bf 58} (1995) 477, hep-ph/9411342.

[6] --- V. F. R. Jones, {\it On Knot Invariants Related to some Statistical Mechanics Models}, 
Pacific J. of Math. {\bf 137} (1989) 311-334. 

[7] --- D. W. L. Sumners, {\it Knot Theory and DNA}, Proc of Symposia in
App. Math.  {\bf 45} (1991) 39-72. 

[8] --- C. B\"auerle, Yu. M. Bunkov, S. N. Fisher, H. Godfrin \&
G. R. Pickett, {\sl Laboratory Simulation of Cosmic String Formation in the 
Early Universe Using Superfluid $^3$He},
Nature {\bf 382} (1995) 332-334. 

[9] --- M. Born \& L. Infeld, {\it Foundations of the New Field Theory}, 
Proc. Roy. Soc. {\bf A144} (1934) 425-451. 

[10] --- Y. Nambu, Lectures at the Copenhagen Summer Symposium  (1970). 

[11] --- B. M. Barbashov \& N. A. Chernikov, {\it Solution and Quantization 
of a 
Nonlinear Two-Dimensional Model for a Born-Infield Type Field},
Soviet Phys. JETP {\bf 23} (1966) 861-868. 

[12] --- S. Brundobler \& V. Elser, {\it Colliding Waves on a Relativistic 
String}, Am. J. Phys. {\bf 60} (1992) 726-732. 

[13] --- B. M. Barbashov \& N. A. Chernikov, {\it Solution of the Two
Plane Wave Scattering Problem in a Nonlinear Scalar Field Theory of the  
 Born-Infield Type},
Soviet Phys. JETP {\bf 24} (1967) 437-442. 

[14] --- M. J. Duff, R. R. Khuri \& J. X. Lu, {\it String Solitons},
Phys. Rept. {\bf 259} (1995) 213, hep-th/9412184 (1994). 

[15] --- P. Goddard, J. Goldstone, C. Rebbi \& C. B. Thorn, {\it Quantum Dynamics
of a Massless Relativistic String}, Nuc. Phys. {\bf B56} (1973) 109-135.

[16] --- J. Scherk, {\it An Introduction to the Theory of Dual Models and Strings},
Rev. Mod. Phys.  {\bf 47} (1975) 123-164. 

[17] --- L. Faddeev \& A. J. Niemi, {\it Stable Knot-Like Structures in
Classical Field Theory}, Nature {\bf 387} (1997) 58-61,
hep-th/9610193 (1996). 

[18] --- M. Zaganescu, {\it Bosonic Knotted Strings and Cobordism Theory}, 
Europhys. Lett. {\bf 4} (1987) 521-525.

\vfill \eject

\centerline{\bf Figure Captions}
\bigskip

{\bf Figure 1:} Oscillations of the simple link   in $M^{2+1}$, calculated from (2.16), for the initial configuration $a^1(\sigma)=\cos(\sigma)$,
$a^2(\sigma)=\sin(\sigma)$ and $b^k(\sigma)=0$, for $k=1,2$. Due to the singularities occurring for $t=\ell /4,3\ell /4,\ldots $, where $\ell=2\pi$ is
the initial configuration length, the circle collapses into a point.  
The world sheet $S$ displayed has parametric representation
$(\cos(t)\cos(p-t),\cos(t)\sin(p-t),t)$,  derived directly from (2.16).

\bigskip

{\bf Figure 2:} Time evolution of the trefoil knot in  $M^{3+1}$, for several
values of $t$ and calculated from (2.16). We represent   
  a tubular neighborhood around the string 
in order to better depict the knot topology.
The world sheet is periodic along the $t$ axis with period $T=\ell/2=18.5942$.
For $T=\ell/4=9.2971$, the trefoil loops collapse and a singularity appears. Singularities occur for $t=(2 n+1)\ell /4$, where $\ell$ is the
string initial length. Due to time parity, the string evolution in
the time interval $\ell/4\le t\le \ell/2$ follows the reverse order
of the time evolution in the interval $0\le t\le \ell/4$.

\bigskip

{\bf Figure 3:} Time evolved configuration of the figure eight   knot
and creation of  singularities in the time interval $0\le t\le \ell/4$.  
\bigskip

{\bf Figure 4:} Zero level set of the functions in (3.3), for the trefoil and
figure eight knot. The dots are the values in the $(p,q)$ plane where singularities of the trefoil knot of Fig. 2 (a) and figure eight knot of
Fig. 3 (b)  occur. The singular $(p,q)$-values correspond to the triple  
intersection of the level lines.   These singularity diagrams are parameterization dependent and correspond to a simple qualitative analysis of
(3.3).

\bye